\documentclass[acmsmall,screen,10pt]{acmart}

\AtBeginDocument{%
  \providecommand\BibTeX{{%
    \normalfont B\kern-0.5em{\scshape i\kern-0.25em b}\kern-0.8em\TeX}}}

\acmJournal{JACM}

\definecolor{BrickRed}{HTML}{B6321C}
\definecolor{ForestGreen}{HTML}{009B55}
\definecolor{Blue}{HTML}{2D2F92}
\definecolor{Red}{HTML}{ED1B23}
\definecolor{Brown}{HTML}{792599}
\definecolor{Purple}{HTML}{99479B}
\colorlet{Black}{black}

{\makeatletter
 \gdef\luismark{%
   \expandafter\ifx\csname @mpargs\endcsname\relax 
     \expandafter\ifx\csname @captype\endcsname\relax 
       \marginpar{\textcolor{red}{Lu\'is~}}
     \else
       \textcolor{red}{Lu\'is~}
     \fi
   \else
     \textcolor{red}{Lu\'is~}
   \fi}
 \gdef\luis{\@ifnextchar[\luis@lab\luis@nolab}
 \long\gdef\luis@lab[#1]#2{{\bf [\luismark \textcolor{red}{#2} ---{\sc #1}]}}
 \long\gdef\luis@nolab#1{{\bf [\luismark \textcolor{red}{#1}]}}
}

{\makeatletter
 \gdef\sidmark{%
   \expandafter\ifx\csname @mpargs\endcsname\relax 
     \expandafter\ifx\csname @captype\endcsname\relax 
       \marginpar{\textcolor{blue}{Sid~}}
     \else
       \textcolor{blue}{Sid~}
     \fi
   \else
     \textcolor{blue}{Sid~}
   \fi}
 \gdef\sid{\@ifnextchar[\sid@lab\sid@nolab}
 \long\gdef\sid@lab[#1]#2{{\bf [\sidmark \textcolor{blue}{#2} ---{\sc #1}]}}
 \long\gdef\sid@nolab#1{{\bf [\sidmark \textcolor{blue}{#1}]}}
}

{\makeatletter
 \gdef\ellenmark{%
   \expandafter\ifx\csname @mpargs\endcsname\relax 
     \expandafter\ifx\csname @captype\endcsname\relax 
       \marginpar{\textcolor{ForestGreen}{Ellen~}}
     \else
       \textcolor{ForestGreen}{Ellen~}
     \fi
   \else
     \textcolor{ForestGreen}{Ellen~}
   \fi}
 \gdef\ellen{\@ifnextchar[\ellen@lab\ellen@nolab}
 \long\gdef\ellen@lab[#1]#2{{\bf [\ellenmark \textcolor{ForestGreen}{#2} ---{\sc #1}]}}
 \long\gdef\ellen@nolab#1{{\bf [\ellenmark \textcolor{ForestGreen}{#1}]}}
}

\usepackage{xspace}
\usepackage{listings}
\usepackage{multirow}

\newcommand{\eg}{{\it e.g.\/,\ }}
\newcommand{\ie}{{\it i.e.\/,\ }}

\newcommand{\js}{JavaScript\xspace}
\newcommand{\dsu}{DSU\xspace}
\newcommand{\mvx}{MVX\xspace}
\newcommand{\domtris}{DOMTRIS\xspace}
\newcommand{\painter}{Painter\xspace}
\newcommand{\nicedit}{nicEdit\xspace}
\newcommand{\colorgame}{Color Game\xspace}
\newcommand{\rrtool}{\texttt{Atbswp}\xspace}

\newcommand{\sysname}{\textsc{Sinatra}\xspace}

\begin{document}

\title{Dynamic Software Updates for Unmodified Browsers through Multi-Version Execution}

\author{Siddhanth Venkateshwaran}
\email{svenka42@uic.edu}
\author{Ellen Kidane}
\email{ekidan2@uic.edu}
\author{Lu\'is Pina}
\email{luispina@uic.edu}
\affiliation{%
  \institution{\\University of Illinois at Chicago}
  \city{Chicago}
  \state{Illinois}
  \country{USA}
}


\renewcommand{\shortauthors}{Venkateshwaran, Kidane, and Pina}

\begin{abstract}

    Browsers are the main way in which most users experience the internet, which makes them a prime target for malicious entities.  The best defense for the common user is to keep their browser always up-to-date, installing updates as soon as they are available.
    Unfortunately, updating a browser is disruptive as it results in loss of user state.
    Even though modern browsers reopen all pages (tabs) after an update to minimize inconvenience, this approach still loses all local user state in each page (\eg contents of unsubmitted forms, including associated \js validation state) and assumes that pages can be refreshed and result in the same contents.
    We believe this is an important barrier that keeps users from updating their browsers as frequently as possible.

    In this paper, we present the design, implementation, and evaluation of \sysname, which supports instantaneous browser updates that do not result in any data loss through a novel Multi-Version eXecution (\mvx) approach for \js programs.
    \sysname works in pure \js, does not require any browser support, thus works on closed-source browsers, and requires trivial changes to each target page, that can be automated.
    First, \sysname captures all the non-determinism available to a \js program (\eg event handlers executed, expired timers, invocations of \texttt{Math.random}).
    Our evaluation shows that \sysname requires 5MB to store such events, and the memory grows at a modest rate of 23.1KB/s as the user keeps interacting with each page.
    When an update becomes available, \sysname transfer the state by re-executing the same set of non-deterministic events on the new browser.
    During this time, which can be as long as 13 seconds, \sysname uses \mvx to allow the user to keep interacting with the old browser.
    Finally, \sysname changes the roles in 353ms, and the user starts interacting with the new browser, effectively performing a browser update with zero downtime and no loss of state.

\end{abstract}



\maketitle

\section{Introduction}

Browsers are the main way in which most users experience the internet.
Browsers are responsible for the safety of user sensitive data, in the form of cookies, saved passwords, and credit card information, and other personal information used to auto-complete forms.
Browsers are also responsible for ensuring the integrity of the websites that the user visits, checking certificates and negotiating encrypted HTTPS channels.
Given all this, browsers are prime targets for malicious entities.
For the common user, the best way to protect their browsers (and the personal data they keep) is to keep the browsers as up-to-date as possible.

Unfortunately, users are slow to update their browsers to a new version.
In terms of percentage of users, data for Google Chrome\footnote{\url{https://gs.statcounter.com/browser-version-market-share/desktop/worldwide/\#daily-20201001-20201201}} and Mozilla Firefox\footnote{\url{https://data.firefox.com/dashboard/user-activity}} show that a new browser version takes about 2 weeks to overtake the previous version, and about 4 weeks to reach its peak.
Given the fast pace of browser releases (6 weeks for Google Chrome and 4 weeks for Mozilla Firefox), the amount of users running outdated versions is significant at any given time.

Browser developers are aware of the problems caused by running outdated versions, and provide features to entice users to update, from reminding the user that a new update is available to minimizing the inconvenience by reopening all pages (tabs) after the update.
Even though popular, the latter feature has two main flaws.
First, it assumes that pages can simply be refreshed after the update.
Such an assumption fails if a login session expires, which causes the page to refresh to the login portal; or if the contents of the page change with each refresh, as is the case with modern social media.
Second, refreshing a page loses all user state accumulated on that page since it was loaded.
Such state includes, among others, data in HTML forms and Javascript state.

The result is simple:  \textbf{Browser updates result in loss of user data}.
We believe that this is an important reason that keeps users from updating their browser, which creates the need for a technique to eliminate data loss due to browser updates.
Dynamic Software Updating (\dsu) techniques can be used for exactly that purpose, updating a program in-process.
Unfortunately, state-of-the-art \dsu tools cannot handle programs as complex as modern commercial internet browsers (Section~\ref{sec:dsu}).
Also, simply dumping the old browser memory state to disk and reloading it in the new browser does not work, as the new browser may change the internal state representation.

Note that the only browser state of interest is the \js state of each page, so we could launch the updated browser and transfer only that state between browsers.
Such an approach has two problems.
First, the user cannot interact with either browser while transferring the state.
Second, the user cannot rollback failed updates, which may result in loss of user data.
Multi-Version eXecution (\mvx) solves both problems by allowing the user to interact with the old browser while the new browser is receiving the state, and by allowing the user to cancel a failed update simply by closing the new browser.
Unfortunately, state-of-the-art \mvx tools cannot handle modern commercial internet browsers (Section~\ref{sec:mvx}), and performing \mvx at the \js is not as straightforward due to the event-driven programming paradigm (Section~\ref{sec:eventloop}).

In this paper, we present the design and implementation of \sysname --- \underline{St}ateful \underline{In}st\underline{a}n\underline{t}aneous b\underline{r}owser upd\underline{a}tes --- a novel \mvx technique implemented in pure \js.
\sysname requires little changes to the target \js application (Section~\ref{sec:design}), which can be performed automatically for all the pages accessed through an HTTP proxy (Section~\ref{sec:arch}).

To perform an update (Section~\ref{sec:phases}), \sysname captures all sources of non-determinism accessed by the browser.
Then, when an update becomes available, \sysname launches the updated browser as a separate process, and feeds it the same non-determinism, thus synchronizing the \js state between both browsers.
At this point, \sysname starts performing \mvx between the two browsers as the user interacts with the outdated browser, which allows the user to build confidence that the update was successful and that no data was lost.
Once the user signals to \sysname that the update was successful, the user can start interacting with the new browser and terminate the old browser.

\sysname captures all sources of non-determinism available to a \js program, including execution of event handlers (Sections~\ref{sec:events}), and non-deterministic functions such as \texttt{Math.random} (Section~\ref{sec:async}).
We implemented \sysname in pure \js using an extra \emph{coordinator} process to enable communication between browsers (Section~\ref{sec:protocol}) that serializes \js non-determinism as JSON (Section~\ref{sec:bubble}).
Our implementation also handles all other sources of state in a \js application (Sections~\ref{sec:bubble} and~\ref{sec:stateful}).

This paper also presents an extensive evaluation of \sysname using 4 \js applications and realistic workloads (Section~\ref{sec:apps}).
Our results show that \sysname runs with very little performance overhead, adding at most 17.38ms to the execution of event handlers (Section~\ref{sec:latency}), which is not noticeable by the user.
For realistic user interactions, \sysname requires 5MB of memory to store the events until a future update happens (Section~\ref{sec:log}).
Furthermore, the amount of memory grows constantly with the length of active user interactions, with a maximum rate of 23.1KB/s (Section~\ref{sec:scale}), which shows that \sysname scales well with typical user interactions with modern websites.


When performing an update, \sysname requires at most 13 seconds to transfer the state between browsers (Section~\ref{sec:catchup}).
We note that the user can continue to interact with the browser during this time.
To switch to the updated browser, \sysname imposes a pause in user interaction of 352ms or less (Section~\ref{sec:promote}), which is perceived as instantaneous.
At its core, \sysname is an \mvx system that delivers events from one browser to another in 55ms or less (Section~\ref{sec:mvxtime}).

In short, this paper has the following contributions:

\begin{enumerate}

    \item The design, and implementation of \sysname, a system for performing \mvx on \js applications.

    \item A technique to use \sysname to perform instantaneous updates to modern commercial closed-source internet browsers, without any loss of state.

    \item An extensive evaluation of \sysname using 4 \js stateful applications.

\end{enumerate}

Upon acceptance, we plan to release \sysname as an open-source project.
We also plan to submit a well-documented artifact that can reproduce the whole evaluation of \sysname.

\section{Background}

Performing \emph{Dynamic Software Updating (\dsu)} on a running browser presents many unique challenges.
First, state-of-the-art \dsu tools require source code changes and do not support programs as complicated as modern internet browsers (Section~\ref{sec:dsu}).
\sysname circumvents that problem by using \emph{Multi-Version eXecution (\mvx)} to perform \dsu~\cite{mvedsua}.
Unfortunately, state-of-the-art \mvx tools also do not support programs as complicated as modern internet browsers (Section~\ref{sec:mvx}).
\sysname moves the level of \mvx from low-level system calls to high-level \js events.
However, performing \mvx in the traditional sense is not possible in \js, due to its event-driven paradigm (Section~\ref{sec:eventloop}).

\subsection{Dynamic Software Updating (\dsu)}
\label{sec:dsu}

Dynamic Software Updating (\dsu) allows to install an update on a running program without terminating it, and without losing any program state (\eg data in memory, open connections, open files).
\dsu has three fundamental problems to solve:  (1) when to stop the running program, (2) how to transform the program state to a representation that is \emph{compatible} with the new version but \emph{equivalent} to the old state, and (3) how to restart the program in the new version.
Solving these problems requires modifying the source code of the updatable program~\cite{rubah,kitsune,mvedsua} adding \emph{safe-update points} to solve problem 1, \emph{state transformation functions} to solve problem 2, and \emph{control-flow migration} to solve problem 3.
This is not possible for popular modern internet browsers (\eg Google Chrome, Microsoft Edge, Apple Safari), as they are closed-source.

Modern \dsu approaches focus on \emph{in-process updates} --- the new version of the program replaces the old version in the same process --- which trivially keep outside resources available between updates (\eg open network connections and files), but limits existing \dsu tools to programs that execute in a single process.
This is not the case of modern internet browsers (\eg Google Chrome uses one process per open tab to improve performance and provide strong isolation between open pages).
Finally, modern internet browsers are examples \emph{self-modifying code} given their Just-In-Time (JIT) \js compiler, which is a well known limitation of state-of-the-art \dsu tools~\cite{kitsune,mvedsua}.
Therefore, existing \dsu tools cannot update modern browsers.


\subsection{Multi-Version eXecution (\mvx)}
\label{sec:mvx}

The main goal of \mvx is to ensure that many program \emph{versions} execute over the same inputs and generate the same outputs.
\mvx can be used to perform \dsu by launching the updated program as a separate process, transferring the state between processes (\eg by forking the original process), and resuming execution on the updated process after terminating the outdated process~\cite{mvedsua}.



Unlike \dsu, state-of-the-art \mvx techniques do not require access to the source code of the target program.
Instead, \mvx interposes \emph{system-calls} through ptrace~\cite{ptrace} or binary-code instrumentation~\cite{varan,freeda}.
This way, \mvx tools can ensure that all processes read the same data, by capturing relevant system-calls (\eg read) and ensuring that they return the same sequence of bytes.


Unfortunately, existing \mvx tools cannot be applied to modern internet browsers.
Doing so results in immediate termination due to \emph{benign divergences} --- equivalent behavior expressed by different sequences of system calls.
For instance, consider how a JIT compiler decides which code to compile/optimize using performance counters based on CPU time.
Interacting with such counters does not result in system-calls, and causes JIT compilers to optimize different code, which then results in different system calls.
It is possible to tolerate such benign divergences~\cite{dsl}, but doing so requires developer support and significant engineering effort, which is not practical.

\mvx also suffers from some of the same issues as \dsu:  no support for multi-process applications, and no support for self-modifying code.

\subsection{\js Messages, Event-Loop, and Non-Determinism}
\label{sec:eventloop}

\begin{figure}[t]
  \centering
  \includegraphics[width=\linewidth]{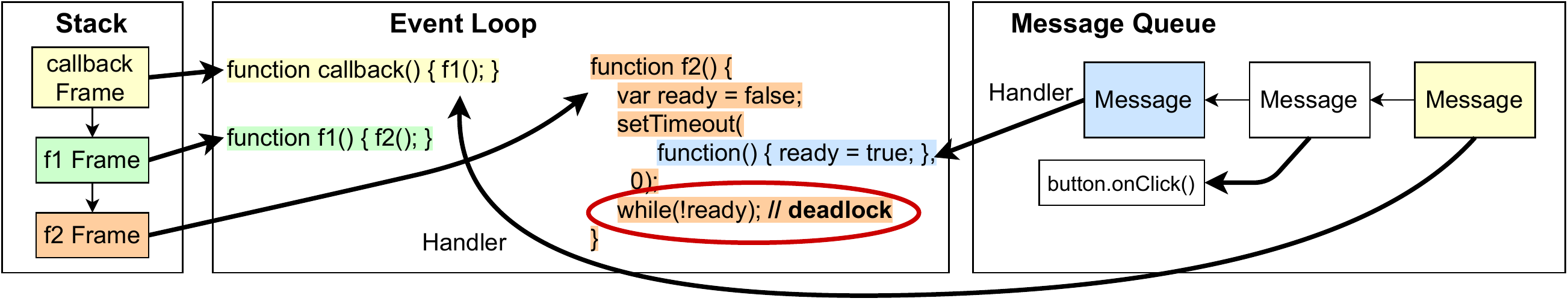}
  \caption{\js's event loop.}
  \label{fig:eventloop}
\end{figure}

\js~\cite{js} is an event-driven programming language animated by an \emph{event loop}, as depicted in Figure~\ref{fig:eventloop}, which processes \emph{messages} from an \emph{event queue}.
The event loop takes one message from the event queue and executes its \emph{handler} to completion.
If the queue is empty, the event-loop simply waits for the next event.
A handler is a \js closure associated with each message.
Given that the event loop is \emph{single-threaded}, there is a single \emph{call stack} and one \emph{program counter} (not depicted) to keep the state of processing the current event.

On a browser, events can come from two sources:  (1) user interaction with DOM elements (\eg \texttt{onclick} on a \texttt{button} element), and (2) browser-generated events, such as expiring timers (\eg \texttt{setTimeout} or \texttt{setInterval}) or receiving replies to pending XML HTTP requests.
Each event generates a message that is added to the end of the event queue.
Besides handlers, messages may have other properties, such as event target (\eg a particular DOM element that generated the message).
Each message corresponds to one \emph{event object} that describes the message in detail (\eg the DOM element and the coordinates of the mouse pointer that triggered a \texttt{mouseover} event).


Events in \js are not executed immediately when triggered, and are instead added to the end of the event queue.
For instance, in Figure~\ref{fig:eventloop}, one button was clicked while the event loop was processing the current message, which results in adding one message to the event queue.

The event-loop is single-threaded and runs each event handler to completion before processing any other event, which has two important consequences.
\textbf{First}, the order and types of events processed are a major part of the non-determinism used to execute a \js program.
Apart from asynchronous non-determinism, described below, rerunning the same events in the same order results in the same execution of the same \js program~\cite{mugshot,dolos,jardis,xcheck,rejs}.
\textbf{Second}, it is not possible for an event handler to issue an event and wait for its completion.
This causes the code in Figure~\ref{fig:eventloop} to \emph{deadlock} when waiting for the flag \texttt{ready} to become \texttt{true}~\cite{stopify} because the handler that sets the flag never executes.
The handler is associated with a timeout (of zero), which adds a message to the end of the queue.
The event loop never finishes executing the current handler, so it never processes any more messages on the queue.

Besides the \emph{synchronous non-determinism} created by events, described above, a \js program can also call functions that are non-deterministic, which we call \emph{asynchronous non-determinism}.
The main non-deterministic functions are \texttt{Math.random}, which generates random numbers between 0 and 1; and methods of the \texttt{Date} object (\eg \texttt{Date.getTime}), which access the current time and date.
Notably, it is not possible to seed the pseudo random number generator behind \texttt{Math.random}.

Given \js's limitations, \textbf{it is not possible to perform traditional \mvx on the asynchronous non-determinism.}
For instance, when generating a random number, typical \mvx approaches ensure each version waits for a message with the same random number (perhaps from a central coordinator process).
In \js's case, this would create the same deadlock as shown above.
Section~\ref{sec:async} describes how \sysname overcomes this limitation.

\section{\sysname Design}
\label{sec:design}

\begin{figure}[t]
\flushleft
{\tt \small \input{sample.html.tex}}
  \caption{Sample HTML code}
  \label{fig:html}
\end{figure}

\sysname supports updating internet browsers through a combination of \mvx and \dsu~\cite{mvedsua}, both at the \js level.
Given that \sysname operates at the \js level, it requires modifications to the pages that an internet browser opens.
Such modifications are shown in Figure~\ref{fig:html}.
Lines~3--6 need to be added, and Line 9 needs to be changed into Line 10.
Adding a \texttt{body.onload} handler ensures that \sysname can execute its logic before any other \js code executes, which is important to intercept all event handlers.
These are quite simple modifications that can be performed automatically by a sophisticated proxy~\cite{mitmproxy} between the browser and the internet, as we describe in Section~\ref{sec:arch}.

After applying the required changes, \sysname leverages the first-class nature of functions in \js, and replaces a number of important functions to intercept all sources of non-determinism:  \texttt{addEventListener}, \texttt{createElement}, \texttt{Math.random}, \texttt{setTimeout}, \texttt{setInterval}, and others.
Sections~\ref{sec:events} through~\ref{sec:mvxjs} describe how \sysname accomplishes this.

\subsection{\sysname Architecture}

\begin{figure}[t]
\centering
\begin{minipage}{.5\textwidth}
  \centering
  \includegraphics[width=\linewidth]{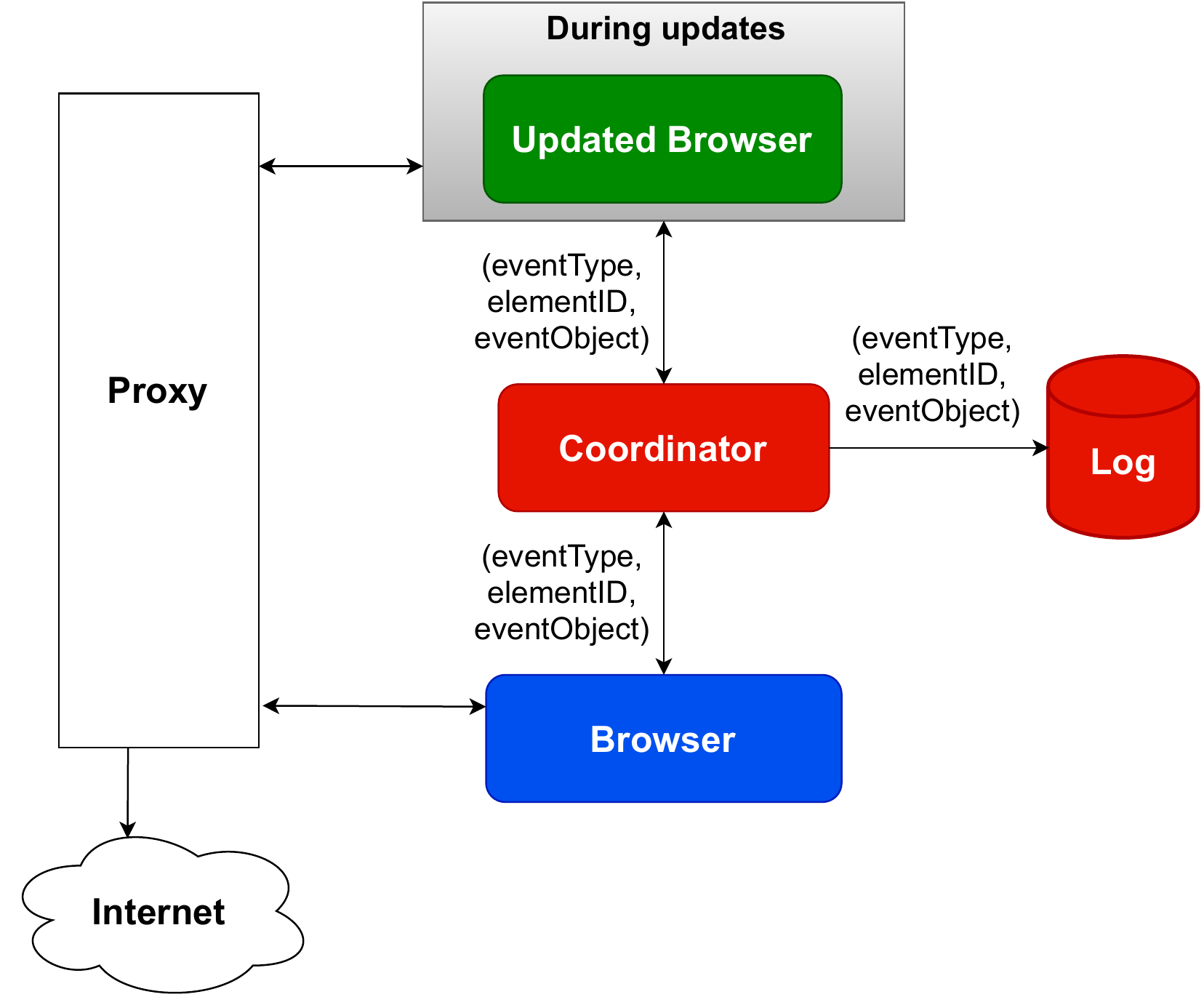}
  \caption{\sysname architecture.}
  \label{fig:arch}
\end{minipage}%
\begin{minipage}{.5\textwidth}
  \centering
  \includegraphics[width=\linewidth]{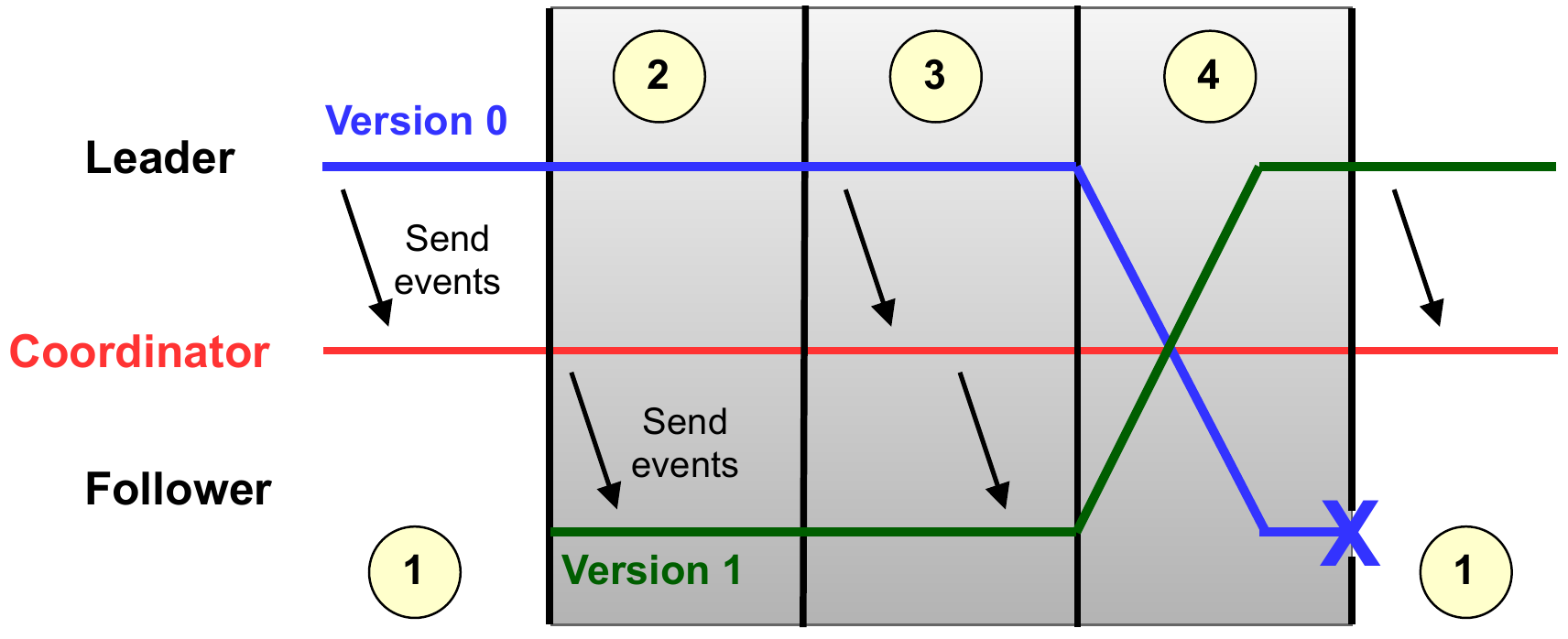}
  \caption{\sysname update phases.
  }
  \label{fig:phases}
\end{minipage}
\end{figure}
\label{sec:arch}

\sysname uses three components at all times, shown in Figure~\ref{fig:arch}: (1) the browser, (2) the coordinator, and (3) a proxy.
When a browser update is available, \sysname requires the new version of the browser to be installed at the same time as the old (current) version.
The updated browser becomes, temporarily, \sysname's fourth component.

Users interact with the browser, which captures \js events and sends them to the coordinator.
The coordinator either saves those events in a log, when there is no update taking place, or sends them to the updated browser, thus performing \mvx.
Users only start interacting with the updated browser after the update is complete, as we describe in Section~\ref{sec:phases}.

The proxy serves three main purposes.
\textbf{First}, the proxy ensures that the updated browser accesses exactly the same content as the original browser did.
Accessing different content means that each browser processes different \js events and leads to benign divergences, as described in Section~\ref{sec:mvx}.
\textbf{Second}, the proxy ensures that outgoing connections remain open while \sysname changes the roles of the browsers.
This is important to ensure that responses to XML HTTP requests are not lost, which can result in errors in the \js application.
\textbf{Third}, as \sysname requires minimal changes to the target page, the proxy performs those changes automatically for all the pages accessed by the browser.
The proxy must also be able to intercept HTTPS traffic.

There is an off-the-shelf proxy that meets all the requirements: \texttt{mitmproxy}~\cite{mitmproxy}.
\texttt{mitmproxy} can intercept HTTPS traffic (through an extra root certificate), is highly configurable with custom Python code, and can redirect traffic from one connection to another.
We validated the feasibility of using \texttt{mitmproxy} for \sysname's purposes through a series of small throw-away prototypes, and we report that \texttt{mitmproxy} can indeed be used with \sysname.
However, for implementation and experimentation simplicity, our current implementation does not use a proxy, as we perform all the changes manually on static HTML pages that do not issue XML HTTP requests.

\subsection{\dsu with \sysname}
\label{sec:phases}

\sysname performs updates over 4 different phases, as shown in Figure~\ref{fig:phases}.

\subsubsection{Phase 1\nopunct} executes for the vast majority of the time, when no update is taking place.
This is the \emph{single-version phase}, which runs a single browser version in isolation.
In this phase, \sysname simply intercepts all \js events and sends them to a \emph{Coordinator} process, which simply keeps them in memory until the events are needed.
Of course, the coordinator needs to have enough memory to store all the events generated on the browser due to user interaction.
Sections~\ref{sec:log} and~\ref{sec:scale} show that \sysname's memory requirements are modest, well within the capabilities of modern computers.
We also note that a webpage simply open (perhaps a tab in the background) generates no events as the user is not interacting with it.

\subsubsection{Phase 2\nopunct} is when updates start, during which the user launches the new browser version.
For each page, \sysname gets all the events from the coordinator and sends them to the new version.
Note that \sysname transfers the state in the background, which allows users to continue to interact with the old browser.
Events generated by user interaction during Phase~2 are simply added to the end of the list of events that the new browser needs to process.
Section~\ref{sec:catchup} shows that \sysname takes, at most, 13 seconds in Phase~2.

\subsubsection{Phase 3\nopunct} starts when the new browser has processed all the events in the coordinator's log.
During Phase~3, \sysname performs \mvx between the old browser and the new browser.
Users interact with the old browser (leader), which sends all events as they happen to the new browser (follower) through the coordinator.
Users can then see the same changes happening in the new browser, side-by-side, as they interact with the old browser, which allows them to verify that \sysname transferred the state correctly, and that the new browser behaves as expected, thus ensuring that the update did in fact work.
If the update did not work, the user can simply terminate the new version and try the update again at a later point.
In this case, \sysname reverts to Phase~1.


\subsubsection{Phase 4\nopunct} starts when the user is confident that the update was successful, and presses a \emph{promote} button on the old browser, which \sysname injects to the top of each page.
\sysname then \emph{demotes} the old browser version, which becomes the follower, and \emph{promotes} the new browser version, which becomes the leader.
The user cannot interact with either browser between pressing the \emph{promote} button and \sysname switching the roles.
This causes the only user-noticeable pause, which we measured in Section~\ref{sec:promote} as 353ms.

From this point on, the user interacts with the new browser version.
\sysname continues \mvx to keep both versions synchronized until the user terminates the old version (now the follower).
At that point, \sysname starts a new \emph{Phase~1}, as the browser was successfully updated with zero downtime and without losing any state.



\subsection{Intercepting Events}
\label{sec:events}

\begin{figure}[t]
  \centering
\flushleft
{\tt \small \input{dom0Interception.js.tex}}
\caption{Interception of a DOM0 event in the leader.  Functions \texttt{sendToCoordinator} and \texttt{registerHandler} shown in Figure~\ref{fig:globalTable} and discussed in Section~\ref{sec:matchels}.}
  \label{fig:dom0Script}
\end{figure}

Through intercepting events and sending them from the leader to the follower (through the coordinator), \sysname establishes the foundation for \mvx and browser updates.
This section explains how \sysname captures browser events in pure \js by intercepting handlers along with their parameters on the leader.

\sysname intercepts events by replacing the original event handler with a special handler.
This way, when a message causes the event loop to execute a handler, \sysname's code executes instead, which allows \sysname to intercept the event that triggered the handler together with the actual handler that is executing.
Messages are generated by either DOM0 or DOM2 event listeners:

\subsubsection{DOM0 events\nopunct}
\label{sec:dom0}
can be registered in-line on the HTML page (\eg Line~12 on Figure~\ref{fig:html}), and through \js properties on the DOM elements (\eg Line~18 on Figure~\ref{fig:html}).
        Event handlers registered at the DOM0 level can be listed by accessing object properties (\eg Lines~2 and~7 on Figure~\ref{fig:dom0Script}).

Intercepting DOM0 events is straightforward, as these handlers can be listed/modified directly from DOM elements, simply by reading/writing the respective property, respectively.
The only challenge is that such events are only present \emph{after} the HTML page loads and executes all in-line scripts.
\sysname's initialization code is the first \js to execute (through the \texttt{body.onload} handler shown in Line~10 of Figure~\ref{fig:html}), which is too early as all elements do not have any DOM0 handler yet.
To wait until the page is loaded, \sysname registers a closure with a zero timeout, which is added to the end of the event queue, as explained in Section~\ref{sec:eventloop}.
When the closure runs, all the page has been loaded, and \sysname can intercept all DOM0 handlers simply by traversing the DOM tree and replacing existing DOM0 handlers with its own special handler.

Figure~\ref{fig:dom0Script} shows the pseudo-code that \sysname uses to intercept DOM0 events.
For each DOM0 handler (Line~1), \sysname starts by capturing the original handler (Line~2) and replaces it with its own closure (Line~6) that captures the current DOM element --- \texttt{element} --- and the event --- \texttt{ev}, sends them to the coordinator (Line~4), and finally runs the handler originally registered by the \js application (Line~5).
Note that \sysname only installs DOM0 events when needed (Line~3).

\subsubsection{DOM2 events\nopunct}
\label{sec:dom2}
register handlers by calling method \texttt{addEventListener} (or \texttt{attachEvent} for Internet Explorer) on the target element (\eg Lines~33--35 on Figure~\ref{fig:html}).
Registering a DOM2 handler requires two arguments.
The first argument specifies the type of event (\eg \texttt{change} for when the target text input box changes).
The second argument is the event handler itself, specified as a \js closure.
Unfortunately, it is not possible to list handlers installed via DOM2.
Furthermore, DOM2 describes a complicated logic about how events ``bubble'' and call all registered event handlers by following the DOM tree and combining DOM0 and DOM2 events.
We discuss how bubbling affects \sysname in Section~\ref{sec:bubble}.

Intercepting DOM2 events is more complicated as it is not possible to list and modify them as explained above for DOM0 handlers.
\sysname intercepts DOM2 events by replacing function \texttt{addEventListener} on all DOM elements with its own closure, as shown in Figure~\ref{fig:dom2Script}.
After the page is loaded, \sysname traverses all DOM elements and, for each element (Line 1), \sysname saves the original \texttt{addEventListener} (Line 2) and replaces it (Line 9).

Note that it is not possible for the underlying \js program to install a DOM2 handler before \sysname installs its own \texttt{addEventListener} because \sysname traverses the DOM tree inside \sysname's \texttt{body.onload} handler, shown in Line~10 of Figure~\ref{fig:html}, which executes before any other \js code.
From this point onwards, when the \js application calls \texttt{addEventListener}, \sysname's code executes instead (Lines 3--7).
To intercept DOM2 events, \sysname installs its own closure using the original \texttt{addEventListener} function (Line 7).
Then, events that trigger the handler execute \sysname's closure which starts by sending the event to the coordinator (Line 4) before calling the original handler that the \js application registered (Line 5).

\begin{figure}[t]
  \centering
\flushleft
{\tt \small \input{dom2Interception.js.tex}}
\caption{Interception of DOM2 events on the leader.  Functions \texttt{sendToCoordinator} and \texttt{registerHandler} shown in Figure~\ref{fig:globalTable} and discussed in Section~\ref{sec:matchels}.}
  \label{fig:dom2Script}
\end{figure}

\subsubsection{Dynamically created elements\nopunct}
\label{sec:dyn}
through function \texttt{document.createElement} have their own \texttt{addEventListener} that escapes \sysname's detection, as such elements are not present when the page is loaded.
\sysname solves this issue by intercepting calls to function \texttt{document.createElement}, in a similar way to how \sysname handles DOM2 events.
First, when the \texttt{body.onload} shown in Line~10 of Figure~\ref{fig:html} executes, \sysname saves the original \texttt{document.createElement} and replaces it with its own.
Then, \sysname can intercept function \texttt{addEventListener} as described in Section~\ref{sec:dom2}, before returning the newly created DOM element to the calling \js code.

Note that \sysname cannot intercept DOM0 events directly.
However, most applications will install DOM0 handlers immediately after getting them from \texttt{document.createElement}.
When intercepting \texttt{document.createElement}, \sysname can add a closure to the message queue, by using a timeout of zero, that will execute before any DOM0 events have a chance to reach a handler of the newly created element.
In that closure, \sysname can intercept all DOM0 events as described in Section~\ref{sec:dom0}.

\subsubsection{Timers and XML HTTP Requests}
\label{sec:timers}
In \js, it is possible to register a closure to execute in the future after a specified time interval, through functions \texttt{setTimeout} --- a one-off event --- and \texttt{setInterval} --- a repeating event, as shown in Lines~3--5 of Figure~\ref{fig:random}.
Of course, such timers are yet another source of synchronous non-determinism that \sysname must handle.
\sysname uses an approach similar explained above in Sections~\ref{sec:dom2} and~\ref{sec:dyn} and replaces functions \texttt{setTimeout} and \texttt{setInterval} with \sysname's own (Line~9).
Then, when the underlying application registers a timer, \sysname transparently intercepts those calls to register its own timer (Lines~13).
When the timer expires, \sysname intercepts the timer event and sends it to the coordinator before executing it (Line~12).

We note that the same approach can be used to intercept XML HTTP Requests, which a \js program can register through function \texttt{new XMLHttpRequest}.
Similarly to intercepting \texttt{addEventListener} in Section~\ref{sec:dom2}, \sysname can intercept function \texttt{XMLHttpRequest.constructor}.
Then, \sysname can use its own function \texttt{send} to list the handlers registered on the relevant properties (\eg \texttt{onreadystatechange}) and replace them with \sysname's own before actually sending the request.
Later, when the XML HTTP Request resolves, the registered \sysname's handlers can perform a logic similar to Lines 4--5 in Figure~\ref{fig:dom2Script}, thus handling this source of non-determinism.
The current prototype of \sysname does not handle XML HTTP requests.

\

\subsection{Intercepting Asynchronous Non-Determinism}
\label{sec:async}

\begin{figure}[t]
  \centering
\flushleft
{\tt \small \input{sample.js.tex}}
\caption{Sample \js program that uses timeouts (Lines~1--5) and \texttt{Math.random} (Line 4); and how \sysname intercepts timer events (Lines~8--14).
    Function \texttt{registerInterval} and \texttt{intervalToCoordinator} are explained Figure~\ref{fig:globalTable} and on Section~\ref{sec:mvxjs}.
}
  \label{fig:random}
\end{figure}

As described in Section~\ref{sec:eventloop}, \js programs can call functions that are non-deterministic.
The most important such function is \texttt{Math.random}, which is used extensively by many \js applications.
Unfortunately, using the same approach described in Section~\ref{sec:events} does not work due to the asynchronous nature of the call to such non-deterministic functions.

For instance, consider the example shown in Figure~\ref{fig:random}.
In this example, there is a list of buttons (Line 2), all disabled.
Every 30 seconds, the program performs an expensive computation (Line 3) and enables one button at random (Line 4).
Every 5 minutes (300 seconds), the program resets all buttons to be disabled (Line 5).

Now consider the following implementation:  \sysname captures the 30 second event, sends it to the coordinator, then captures the execution of \texttt{Math.random}, and also sends it to the coordinator.
This approach works if all the events are known in-advance (\ie Phase 2 of Figure~\ref{fig:phases} and existing record-replay approaches).
However, \textbf{this approach does not work for \mvx} (\ie Phase 3 and 4 of Figure~\ref{fig:phases}).
In this case, it is possible that the follower receives the timer event and reaches the call to \texttt{Math.random} before the leader, as the time required to perform the expensive computation may not match in both versions, and the follower may complete it before the leader due to OS scheduling, or simply different performance between browsers.
At this point, the follower does not know which number to return to match the leader.
Making matters even worse, the follower cannot simply wait for the leader, because doing so in \js's event-loop model results in a deadlock, as explained in Section~\ref{sec:eventloop}.

Functions that result in asynchronous non-determinism thus need special consideration.
To deal with \texttt{Math.random}, \sysname starts by generating a sequence of $N$ random numbers when the page loads (\ie in the \texttt{body.onload} function on Line~10 of Figure~\ref{fig:html}).
Then, each call to \texttt{Math.random} consumes one number from the sequence.
The leader replenishes the sequence by sending a fresh random number to the coordinator for each number consumed.
This approach allows a fast follower to consume up to $N$ random numbers asynchronously at its own pace, ensures all random numbers match between leader and follower, and provides a fresh supply of browser-grade randomness.
In practice, we found that a cautious value of $N=100$ works well.

Other functions can be sources of asynchronous non-determinism, such as \texttt{Date} methods.
A similar approach can be used, in which there is a buffer of $N$ dates with $100ms$ of difference, which allows the follower to consume dates at its own pace, the leader to replenish them as it can, and both browsers to get the same results.
Given the different nature of time, this approach can be supplemented with each message sent to the coordinator having its own timestamp, which is fed directly into the buffer of times.
In practice, however, we found that \texttt{Date} divergences simply lead to slightly different content in each browser as \texttt{Date} results are simply shown to the user and not used in other computations, and the current version of \sysname does not handle \texttt{Date} methods.



\subsection{Multi-Version Execution in \js}
\label{sec:mvxjs}

So far, this document describes how to capture all the sources of non-determinism used by a \js program on the leader browser.
But this is only one half of the problem.
To transfer the state between browsers, and to keep them synchronized after that, \sysname needs to ensure that the follower browser sees exactly the same non-determinism (\ie the same events in the same order on the same DOM elements).

\subsubsection{Read-only Follower}
In the context of \mvx we now have two browsers, as shown in Figure~\ref{fig:phases}.
The user interacts with a \emph{leader} browser, which sends all the non-determinism to the coordinator process.
Then, a \emph{follower} browser receives the same non-determinism from the coordinator.
This way, \sysname ensures both leader and follower are always synchronized.

Users can inspect the state of the follower browser, but they cannot modify it.
To do so, the follower also intercepts all the handlers as described in Section~\ref{sec:events}, but does not install any event handlers with the browser.
Instead, the follower registers just with \sysname (\ie Line~7 in Figure~\ref{fig:dom0Script} sets \texttt{onclick} to \texttt{null}, Line~7 in Figure~\ref{fig:dom2Script} and Line~13 in Figure~\ref{fig:random} are omitted).
This way, users cannot change the state on the follower.
Also, this ensures that the follower executes timer handlers in sync with the leader, running them only when the leader sends the respective events.

\subsubsection{Matching elements}
\label{sec:matchels}

\begin{figure}[t]
    \flushleft
{\tt \small \input{globalHandlerTable.js.tex}}
\caption{Matching events and timers to handlers and elements in both leader and follower browsers.}
  \label{fig:globalTable}
\end{figure}

\sysname assigns IDs (monotonically increasing numbers) to each DOM element that does not already have an ID when the page is loaded, during the \texttt{body.onload} handler (shown in Line~10 of Figure~\ref{fig:html}).
\sysname does the same for dynamically added elements (described in Section~\ref{sec:dyn}).
Given that \sysname traverses the same DOM tree in a deterministic way, and executes \texttt{createElement} in the same order in both browsers, the same element always receives the same ID in both browsers.

\sysname keeps a global structure with all the handlers registered, as shown in Figure~\ref{fig:globalTable} (Line~2).
When registering events, \sysname keeps a map for each element ID (Lines~7 and~8).
The map associates event types (\eg \texttt{onclick}) to the respective handler (Line~10) and the target element in which the event was registered (Line~11).
We note that each browser keeps references to its own handler and element.

The leader sends events to the coordinator via function \texttt{sendToCoordinator}, which serializes the event as discussed in Section~\ref{sec:bubble}.
The follower receives events from the coordinator via function \texttt{receiveFromCoordinator}, which consults the global structure to get the target element (Line~25) and the handler (Line~26) registered for the current event being triggered.
Then, the follower calls the handler directly, setting the receiver as the target element (Line~27).
Note that the event is deserialized before calling \texttt{receiveFromCoordinator}, as described in Section~\ref{sec:bubble}.


\subsubsection{Matching timers}
For timers and XML HTTP Requests, described in Section~\ref{sec:timers}, the follower never registers these with the browser.
Instead, the follower executes the closures registered with each handler in the order that the leader issues them through the coordinator.
However, this creates a problem:  How can the follower distinguish between many different closures?
For instance, consider the example shown in Figure~\ref{fig:random}.
This example installs two closures associated with different timeouts, one in Line~3 and another in Line~5.
When one of these expires and the leader executes it, how can the follower know which to execute?

First, \sysname hashes the code text (\ie the source) of each closure registered with a timer in both leader and follower (Line~10 in Figure~\ref{fig:random}).
Given that both register the same handlers, the hash matches between versions.
Both versions then keep a table from hashes to closures and delays (Line~3 and Lines~13--17 in Figure~\ref{fig:globalTable}).
When sending an event about an expired timer, the leader sends the hash of the closure associated with the timer (Line~12 in Figure~\ref{fig:random}).
The follower then uses the hash to address its table, get the correct closure, and execute it (Line~29 in Figure~\ref{fig:globalTable}).

\subsubsection{Promotion/demotion}\label{sec:swap}
When the roles of the browsers switch (Phase 3 in Figure~\ref{fig:phases}), \sysname uses the information kept on the global structure (Line~2 on Figure~\ref{fig:globalTable}) to installs all event handlers on the respective DOM elements on the promoted browser, and removes them on the demoted browser.
\sysname also cancels all the timers on the demoted browser, and installs them on the promoted browser (with the original timeout value).
Because \sysname does not track how much time passed since each timer was installed, this approach may cause timers to expire after longer than needed ($2\times$ in the worst case).
However, this is correct as timers in \js guarantee only a minimum amount of time to wait before triggering the associated closure.

\section{Implementation}
\label{sec:implementation}

In this section, we describe the implementation details of \sysname.
\sysname is implemented in pure \js, totaling \texttt{1127} lines of code.
The web APIs leveraged by \sysname to intercept user and system generated events are compatible with the most recent versions popular browsers, such as Google Chrome, Mozilla Firefox, Apple Safari, and others.
\sysname works out of the box for most browsers, without requiring external packages, tools, or plugins.

\subsection{Coordinator and Protocol}
\label{sec:protocol}

The coordinator process enables communication between both browsers, which is at the core of \sysname's approach to \mvx, and keeps a log of \js events during Phase~1, as shown in Figure~\ref{fig:phases}.
We implemented the coordinator process using \emph{node.js}~\cite{nodejs}, so it executes in its own separate (headless) process without a browser.
We use the \emph{SocketIO}~\cite{socketio} \js library to enable bi-directional communication between the coordinator and each browser.

The initialization protocol for \sysname is quite simple.
First, the coordinator should be executing before any browser is launched.
On browser launch, \sysname starts by connecting to the coordinator using a pre-configured address and port, and sends a message.
The coordinator replies with the role of this browser, which is \emph{leader} for the first browser and \emph{follower} for the second.

Upon learning its role, a leader browser generates the list of random numbers mentioned in Section~\ref{sec:async}, sends it to the coordinator, and starts sending all events from that point on.
A follower browser, conversely, waits for the coordinator to send the list of random numbers, followed by the events that were kept during the leader's execution.
At this point, the coordinator informs the follower that it is synchronized, and \mvx starts.
During \mvx, the coordinator sends each event, received from the leader, to the follower as it is received.

Given that communication is bidirectional, the coordinator does not have to establish new channels when promoting the follower/demoting the leader.
Instead, each browser simply changes roles and execution continues in \mvx, but in the opposite direction.
The promotion/demotion event starts from the leader (outdated browser), when the user presses a button that \sysname injects at the top of the page.
Together with a special promotion/demotion message, \sysname also sends the list of pending timers that were cancelled and their timeouts, as described in Section~\ref{sec:swap}.

\subsection{Serializing/Deserializing Events and Bubbling}
\label{sec:bubble}

When the underlying \js program executes an event handler on the leader, \sysname's code is first called with the event.
First, \sysname gets the \textbf{name} of the event, (\eg defined as argument \texttt{eventType} in Line~3 of Figure~\ref{fig:dom2Script}).
Second, \sysname gets the \textbf{ID of the target element} --- the element that triggered the change (\eg defined as argument \texttt{element.id} in Line~3 of Figure~\ref{fig:dom2Script}).
As explained in Section~\ref{sec:matchels}, \sysname ensures that all elements have a unique ID.
Finally, \sysname creates a \js object to hold a copy of the event object, and populates it with \textbf{all the fields in the event object}, which include the coordinates of mouse events, which key was pressed that triggered the event, and other relevant data.

At this point, \sysname can send the \js object to the coordinator.
The SocketIO implementation automatically turns the \js object into its JSON representation~\cite{json} through function \texttt{JSON.stringify} on send, and back into a \js object using function \texttt{JSON.parse}.
The coordinator simply keeps a list of tuples (name, element ID, event) received from the leader.
Sending this list to the follower, when it becomes available, requires another round of serializing to JSON by the coordinator, and deserializing back into \js objects by the follower.

An important note is that \sysname feeds the deserialized event directly into each handler in the follower, as shown in Line~27 of Figure~\ref{fig:globalTable}.
\sysname does not create/trigger a new synthetic browser event, as some record-replay systems for \js do through \texttt{DOMnode.fireEvent}~\cite{mugshot}.
This design decision simplifies handling \emph{event bubbling}, when many handlers trigger for a single event (\eg when a child DOM element has a different handler for the same event as its parent).
Instead, \sysname simply captures the order in which the leader executes the event handlers, and their respective targets; and then calls the same handlers by the same order in the follower.
The alternative of creating synthetic events has well-known corner-cases that require special consideration.
Furthermore, \sysname can handle browser updates that change the bubbling behavior.

\subsection{Stateful DOM Elements and Text Selection}
\label{sec:stateful}

DOM elements, such as radio buttons, check boxes, and text boxes, keep internal state.
For instance, when the user selects a check box, the state of that check box changes (it is now selected).
Updating the state does not execute any \js handler, which means that \sysname cannot intercept it directly.
Fortunately, there are only a limited number of such elements, and \sysname handles them as a special case by installing its own event handler associated with the \texttt{change} event, even when there is no application handler.
The event handler simply captures the updated state of the DOM element, which allows the follower to remain synchronized with the execution on the leader.

Another source of state, and non-determinism, is the text selected by the user.
\js can access and manipulate the current text selection, based on a range of characters on a text element (start and end).
To detect when the text selection changes, the leader listens for left mouse button releases, and \texttt{SHIFT} key releases.
At that point, \sysname can obtain the current text selection (if any), create a \js object that captures the start and span of the selection, and send it to the coordinator.
On the follower side, \sysname uses the information received to select the same text.

\section{Experimental Evaluation}

In this section, we evaluate the feasibility of using \sysname to deploy browser updates in practice, by measring the overhead it introduces in terms of perceived latency by the user, and extra memory needed by the user's computer.
We also evaluate \sysname's performance as an \mvx tool to enable future research.
In that regard, we pose the following \emph{research questions (RQs)}:

\begin{itemize}

    \item \textbf{RQ1:}  Is the latency added by \sysname noticeable by the user?

    \item \textbf{RQ2:}  What is the average size of the log that \sysname keeps?

    \item \textbf{RQ3:}  How does the size of the log that \sysname keeps grow with user interaction?

    \item \textbf{RQ4:}  How long does \sysname take to perform a browser update?

    \item \textbf{RQ5:}  What is the latency when \sysname is used as a \js \mvx system?

\end{itemize}

We used two versions of two popular internet browsers, Mozilla Firefox versions \texttt{82.0} and \texttt{83.0}, and Google Chrome versions \texttt{88.0.4323.150} and \texttt{89.0.4389.72}.
Unless when using updates, both leader and follower used the lowest version of each browser.
The experimental evaluation took place in two modern laptop computers running Ubuntu Linux 20.04 LTS 64bit: (1) an Intel(R) Core(TM) i7-7500U CPU 2.70GHz with 8GB of RAM, and (2) an Intel(R) Core(TM) i7-8565U CPU 1.80GHz with 12GB of memory.

\subsection{Applications and Workloads}
\label{sec:apps}

We evaluated \sysname with the 4 \js applications (describe below).
Each application requires user interaction, using the keyboard and/or mouse.
We automated such interaction using the tool \rrtool~\cite{atbswp} to record mouse and keyboard interactions --- \emph{workloads} --- for each application, and then replay them.
\rrtool records mouse and keyboard interactions and writes an executable Python script that replays those events using the library \texttt{pyautogui}~\cite{pyautogui}.
We now describe each program, and the workload we used:

\subsubsection{\nicedit~\cite{nicedit}\nopunct}
uses \js to add a rich-text editing toolbar to an HTML \texttt{div} element.
        The toolbar applies styles (\eg bold, italic, underline, font, color, size) to the text selected via the \texttt{document.execCommand} \js API~\cite{execommand}.
        \nicedit creates the toolbar dynamically, using \texttt{document.createElement} to generate buttons and custom screens (\eg to input the URL and text of an hyperlink), and attaches DOM0 event listeners to each generated element (\ie buttons on the toolbar).
        \nicedit also creates a \texttt{textarea} element dynamically, where the user can input text.

        The workload for \nicedit starts with a pre-generated text, selects sections of text, and edits each in a different way: making the text bold, italic; changing the font size, font family (Arial, Helvetica, etc), and font format (heading and paragraph).
        The workload also changes the indentation of a paragraph, increasing it twice and then returning the paragraph back to its original indentation.
        This workload is representative of a user editing some text in a document, and was performed using Machine~1.

\subsubsection{\domtris\cite{domtris}\nopunct}
is a dynamic-HTML based Tetris game that uses \js to implement the game mechanics:  generate random pieces of different sizes, shapes, and colors; intercept user input; and schedule the movement of each piece inside the Tetris board.
        \domtris uses \texttt{setTimeout} to schedule the downwards movement of the current piece, and \texttt{Math.random} to pick the next piece from the available set of pieces.
        The player interacts with \domtris solely via the directional arrows on the keyboard, intercepted via DOM2 event listeners.
        Each piece is created with \texttt{document.createElement}.

        We automated a Tetris game that drops each piece (without rotating it) on the center, extreme left, and extreme right of the board.
        As time goes on, the board fills along the center and edges until there is no space left, at which point the game ends.
        We understand that this is not how people play Tetris, but we cannot use \rrtool to automate a valid Tetris game because \domtris uses a different random seed for every game.\footnote{We could use \sysname's interception of \texttt{Math.random} to generate a fixed sequence of random values.  However, this would not show that \sysname keeps the random values synchronized between browsers.}
        This workload used Machine 2.

\subsubsection{\painter\cite{painter}\nopunct}
allows users to draw pictures using the mouse, with various colors and tools (free-hand brushes, lines, rectangles and circles).
        The user interacts with \painter using only the mouse over 3 HTML5 \texttt{canvas} elements~\cite{canvas}: (1) the tool set, (2) the drawing area, and (3) the color and line-width picker.
        \painter tracks the mouse position and button click/drag using DOM2 events, and reacts to different tool and color selections using DOM0 events.
        Given that \painter tracks the mouse movements at all times, it generates a large number of events.

        Our workload draws a tic-tac-toe board with the brush and line tools, then draws different shapes of different colors inside the board.
        This requires selecting different tools, colors, and brush strokes; effectively interacting with all parts of \painter.
        Note that \rrtool records the mouse with coarse precision between mouse clicks, which results in a low fidelity replay.
        For instance, when dragging the mouse along a line, \rrtool only captures the mouse position on the start of the line when the mouse button is pressed, one or two positions along the line, and the final position when the mouse button is released.
        We edited the generated Python script to ensure that the recording replays mouse movements on a pixel-by-pixel basis, to ensure high fidelity and accurate event counts.
        Unfortunately, due to \texttt{pyautogui}'s low performance when replaying a large number of mouse movements, the \painter workload takes minutes to execute what took us seconds to draw.
        This workload used Machine 1.

\subsubsection{\colorgame\cite{colorgame}\nopunct}
is a game that tests reaction time via the Stroop Effect~\cite{stroop} (delay in reaction time between congruent and incongruent stimuli).
        The game shows players one color name, and requires players to press the button with the same name (out of 4 buttons), but with a different background color (\eg press the red button with text ``Blue'' when the game specifies the color ``Blue'').
        The game keeps track of the score ($+5$ for each correct click, $-3$ otherwise) during a 30 second round.
        Interestingly, \colorgame uses the Angular JS framework~\cite{angular} to draw all elements.
        Internally, Angular uses \texttt{document.createElement}, a combination of DOM0 and DOM2 event handlers, \texttt{setTimeout}, \texttt{setInterval}, and \texttt{Math.random}.

        The workload consists of one run of the game (lasting 30 seconds).
        The workflow consists of clicking each of the four color buttons in arbitrary order after every one second until the game ends.
        This workload used Machine 2.

\subsection{\sysname Latency}
\label{sec:latency}

\begin{table}[t]
    \centering
    \caption{Time required to run event handlers, average of 5 runs with standard deviation.}
    \begin{tabular}{| l | l | c | c | c | c |}
        \hline
        \multirow{2}{*}{\textbf{Program}} & \multirow{2}{*}{\textbf{Browser}} &\multirow{2}{*}{\textbf{Vanilla}} & \multirow{2}{*}{\textbf{\sysname}} & \multicolumn{2}{c|}{\textbf{Overhead}} \\
        \cline{5-6}
        & & & & \textbf{Relative} & \textbf{Absolute} \\
        \hline
\multirow{2}{*}{\nicedit} & Firefox	& $	3.097	ms \pm	0.480	$ & $	10.682	ms \pm	2.085	$ & $	3.449	\times $ & $+	7.585	ms$ \\
~ & Chrome	& $	8.714	ms \pm	0.840	$ & $	26.094	ms \pm	4.561	$ & $	2.995	\times $ & $+	17.380	ms$ \\
\hline													
\multirow{2}{*}{\painter} & Firefox	& $	0.123	ms \pm	0.006	$ & $	3.686	ms \pm	0.248	$ & $	29.969	\times $ & $+	3.563	ms$ \\
~ & Chrome	& $	0.074	ms \pm	0.017	$ & $	5.452	ms \pm	0.248	$ & $	73.497	\times $ & $+	5.378	ms$ \\
\hline													
\multirow{2}{*}{\domtris} & Firefox	& $	0.685	ms \pm	0.092	$ & $	1.417	ms \pm	0.062	$ & $	2.069	\times $ & $+	0.732	ms$ \\
~ & Chrome	& $	1.032	ms \pm	0.027	$ & $	1.707	ms \pm	0.093	$ & $	1.654	\times $ & $+	0.675	ms$ \\
\hline													
\multirow{2}{*}{\colorgame} & Firefox	& $	0.643	ms \pm	0.041	$ & $	1.984	ms \pm	0.081	$ & $	3.085	\times $ & $+	1.341	ms$ \\
~ & Chrome	& $	0.697	ms \pm	0.005	$ & $	4.042	ms \pm	0.162	$ & $	5.801	\times $ & $+	3.346	ms$ \\

        \hline
    \end{tabular}
    \label{tab:latency}
\end{table}

To measure the extra latency added by \sysname to each event on the leader, we compared the execution of each program without \sysname (\emph{vanilla}) and with \sysname.
The vanilla version measures the time taken to execute each original event handler during the workload.
The \sysname version measures the time taken to also execute \sysname's logic together with the original event handler.
We measure the runtime of each event handler triggered during the workload, and report the average time among all the event handler executions observed.
Table~\ref{tab:latency} shows the results.


This experiment highlights the extra latency that \sysname adds to each event.
Table~\ref{tab:latency} shows that \sysname increases the latency by a maximum of $+17ms$ (\nicedit on Chrome), from $8ms$ to $26ms$.
The maximum increase in relative terms is for \painter on Chrome, at $73\times$, which translates to a low absolute increase from $0.07ms$ to $5ms$.
The results answer \textbf{RQ1}: \textbf{Users cannot notice the extra latency introduced by \sysname}.

\subsection{Log sizes}
\label{sec:log}
\sysname spends the vast majority of the time executing in single-leader mode, as described in Section~\ref{sec:phases}.
In this mode, \sysname stores a log in the coordinator with all the events and handlers that the (single) leader executed.
In this experiment, we executed the workload for each application in single-leader mode to measure the size of the log on the coordinator, in number of events and size of the log.
Table~\ref{tab:size} shows the results.

We can see that the number of events varies widely between different experiments.
\nicedit has the smallest number of events, as styling text results in a low number of button clicks and text selections.
\domtris and \colorgame have a similar number of events, which involve user input, timers expiring, and random number generation.
Finally, as expected, \painter generates the largest number of events due to its fine-grained tracking of mouse events.
In terms of absolute log size, we can see that all logs are below 5.5MB.
The results of this experiment allow to answer \textbf{RQ2}:  \textbf{\sysname requires a modest amount of memory to store the log, below 5.5MB per page}.
This result shows the practical applicability of \sysname, given that average modern computers measure memory in tens of GB.

\begin{table}[t]
    \centering
    \caption{Log sizes in single-leader mode, average of 5 runs with standard deviation.}
    \begin{tabular}{| l | l | c | c |}
        \hline
        \textbf{Program} & \textbf{Browser} & \textbf{\# of events} & \textbf{Log size (bytes)} \\
        \hline
\multirow{2}{*}{\nicedit} & Firefox	& $	74	\pm	1.34	$ & $	145,034	\pm	2,306	$ \\
~ & Chrome	& $	79	\pm	0.00	$ & $	137,639	\pm	415	$ \\
\hline									
\multirow{2}{*}{\painter} & Firefox	& $	2,055	\pm	13.63	$ & $	5,553,246	\pm	37,680	$ \\
~ & Chrome	& $	2,042	\pm	9.13	$ & $	5,052,129	\pm	24,433	$ \\
\hline									
\multirow{2}{*}{\domtris} & Firefox	& $	549	\pm	15.34	$ & $	1,104,293	\pm	2,527	$ \\
~ & Chrome	& $	544	\pm	14.65	$ & $	1,070,255	\pm	2,446	$ \\
\hline									
\multirow{2}{*}{Color Game} & Firefox	& $	635	\pm	0.00	$ & $	400,857	\pm	540	$ \\
~ & Chrome	& $	638	\pm	1.34	$ & $	482,002	\pm	1,609	$ \\
        \hline
    \end{tabular}
    \label{tab:size}
\end{table}

\subsection{\sysname scalability}
\label{sec:scale}

User interactions with websites may differ in length of time and number of events triggered.
To measure how \sysname behaves with different lengths of interaction, we designed an experiment that uses 3 modified workloads for each application ---  small, medium, and large --- modified as follows.

\begin{description}

    \item[\nicedit] Repeat the experiment $N$ times, each time performing the same various text changes that have been described previously.
	  \textbf{Small:}  $N=2$.
        \textbf{Medium:} $N=4$.
        \textbf{Large:}  $N=6$.

    \item[\domtris] Move Tetris pieces to one, two, or three sides of the board.
        \textbf{Small:}    Left side only.
        \textbf{Medium:}  Left and right sides.
        \textbf{Large:}    Left and right sides, and middle of the board.

    \item[\painter] Repeat the drawing $N$ times, pressing the \emph{``Clear''} button (which clears the canvas) in between.
        \textbf{Small:}  $N=2$.
        \textbf{Medium:} $N=4$.
        \textbf{Large:}  $N=6$.

    \item[\colorgame] Play a game $N$ times, restarting it at the end of each 30 second run by pressing the \emph{``Restart''} button.
        \textbf{Small:} $N=1$.
        \textbf{Medium:} $N=3$.
        \textbf{Large:} $N=5$.
    
\end{description}

We repeated the experiment for each size, in single-leader mode, and we measured the duration (seconds), the total number of events in the log (thousands), the size of the log (MB), and the bandwidth needed to send all the events (KB/s).
The bandwidth is not measured directly, but computed from the duration and the size of the log, and it shows how much the log grows as the user keeps interacting with a page over time.
Table~\ref{tab:growth} shows the results.

For most experiments, the bandwidth remains about constant even as the length of interaction increases, which is to be expected.
\colorgame is the notable exception, in which the bandwidth increases with the length (and intensity) of user interaction.
We believe this is due to internal AngularJS behavior that: (1) never cancels timers with the browser, simply executes a test to return from cancelled timers, which results in more timers expiring as the game is played again and again; and (2) installs \texttt{hover} handlers for all elements, which call \texttt{Math.random}, which result in more handlers executing as the experiment moves the mouse to the \emph{``Replay''} button, and back to the playing area.
\painter results in the largest log files, and the highest bandwidth, because it targets all the mouse movements with a fine level of detail (pixel by pixel, as discussed above).
Still, the log bandwidth stays under 21.5KB/s, which is acceptable.

We note that the original \painter interaction took about 20sec, the runtimes shown in Table~\ref{tab:growth} are artificially inflated by the slow speed of \texttt{pyautogui}.
Still, the original bandwidth would be around 1MB/s, which is still acceptable for applications that track the mouse with fine detail.

The results of this experiment provide an answer to \textbf{RQ3}: \textbf{\sysname logs grow at a rate of 23.5KB/s as the user interacts with the page.}
This result is acceptable, as mouse-based user interactions are short and the bandwidth is not a bottle-neck for inter-process communication.
We note that the result is much smaller for all the other cases.

\begin{table}[t]
    \centering
    \caption{Duration, number of events, log size, and bandwidth needed for growing workloads. \textbf{S} means \emph{small}, \textbf{M} means \emph{medium}, and \textbf{L} means \emph{large}.  \textbf{FF} means \emph{Firefox}, and \textbf{Chr} means \emph{Chrome}.  Average of 5 runs.}
    \begin{tabular}{| l | l | c | c | c | c | c | c | c | c | c | c | c | c |}
        \hline
        ~ & ~
                             & \multicolumn{3}{c|}{\textbf{\nicedit}}
                             & \multicolumn{3}{c|}{\textbf{\domtris}}
                             & \multicolumn{3}{c|}{\textbf{\painter}}
                             & \multicolumn{3}{c|}{\textbf{\colorgame}} \\
\cline{3-14}
                             ~ &
                             & S & M & L
                             & S & M & L
                             & S & M & L
                             & S & M & L \\
        \hline
        \hline

\textbf{Duration} & FF & $	72	$ & $	106	$ & $	140	$ & $	76	$ & $	96	$ & $	113	$ & $	500	$ & $	963	$ & $	1,427	$ & $	65	$ & $	143	$ & $	221	$ \\
\cline{2-14}																								
\textbf{(sec)} & Chr & $	69	$ & $	103	$ & $	137	$ & $	76	$ & $	96	$ & $	113	$ & $	492	$ & $	948	$ & $	1,407	$ & $	75	$ & $	152	$ & $	229	$ \\
\hline																								
\hline																								
\textbf{\# of events} & FF & $	0.10	$ & $	0.24	$ & $	0.36	$ & $	0.30	$ & $	0.43	$ & $	0.52	$ & $	4.12	$ & $	8.25	$ & $	12.40	$ & $	0.60	$ & $	2.52	$ & $	5.40	$ \\
\cline{2-14}																								
$\times 1000$ & Chr & $	0.12	$ & $	0.23	$ & $	0.34	$ & $	0.30	$ & $	0.42	$ & $	0.52	$ & $	4.09	$ & $	8.19	$ & $	12.33	$ & $	0.60	$ & $	2.52	$ & $	5.40	$ \\
\hline																								
\hline																								
\textbf{Size of log} & FF & $	0.28	$ & $	0.56	$ & $	0.84	$ & $	0.62	$ & $	0.87	$ & $	1.08	$ & $	11.09	$ & $	22.27	$ & $	33.46	$ & $	0.35	$ & $	2.01	$ & $	4.96	$ \\
\cline{2-14}																								
\textbf{(MB)} & Chr & $	0.26	$ & $	0.52	$ & $	0.77	$ & $	0.60	$ & $	0.85	$ & $	1.05	$ & $	10.12	$ & $	20.34	$ & $	30.55	$ & $	0.32	$ & $	1.85	$ & $	4.56	$ \\
\hline																								
\hline																								
\textbf{Bandwidth} & FF & $	3.9	$ & $	5.3	$ & $	6.0	$ & $	8.2	$ & $	9.1	$ & $	9.6	$ & $	22.2	$ & $	23.1	$ & $	23.5	$ & $	5.4	$ & $	14.0	$ & $	22.4	$ \\
\cline{2-14}																								
\textbf{(KB/s)} & Chr & $	3.8	$ & $	5.0	$ & $	5.6	$ & $	7.9	$ & $	8.8	$ & $	9.3	$ & $	20.6	$ & $	21.5	$ & $	21.7	$ & $	4.3	$ & $	12.2	$ & $	19.9	$ \\

        \hline
    \end{tabular}
    \label{tab:growth}
\end{table}

\subsection{Browser updates with \sysname}

\sysname can be used to deploy a browser update without incurring any loss of (\js) state on the pages opened by the running browser.
Such updates are multi-step processes, in which (1) the updated browser, running as follower, processes the log it receives from the coordinator to synchronize its state with the leader; and (2) the follower is promoted to be the new leader.
This section describes two experiments, one for each of the steps.

\subsubsection{Log processing time}
\label{sec:catchup}

This experiment measures the time that the updated browser, running as follower, takes to process all the events in the log sent by the coordinator.
We executed the workload for each program (to completion), and then launched the new browser as a follower.
On the follower, we took two measurements: (1) the time taken since the page is loaded until the follower is up-to-date with the leader, and (2) the time taken just processing the event log sent by the coordinator.
Note that (1) includes all the \sysname initialization logic plus (2).
Columns ``Process log'' and ``Start executing'' of Table~\ref{tab:update} show the results for (2) and (1), respectively.

We can see that processing the log of events is an important portion of the overall time required to start a follower.
Most cases take just under one second, except \nicedit and \painter for Chrome.
\nicedit takes a comparable time in Chrome and Firefox.
\painter takes much longer to process the events in Chrome (11 seconds vs 250ms).
We believe this is due to internal performance differences in deserializing JSON into \js objects between the two browsers, given the sheer difference in log size when compared with other experiments.
We are still investigating the cause of this discrepancy.

\begin{table}[t]
    \centering
    \caption{Time from launching a follower until its state is up-to-date.  Average of 5 runs with standard deviation.}
    \begin{tabular}{| l | l | c | c | c |}
        \hline
        \textbf{Program} & \textbf{Browser} & \textbf{Process log (ms)} & \textbf{Start executing (ms)} & \textbf{Promote (ms)} \\
        \hline
\multirow{2}{*}{\nicedit} & Firefox	& $	353	\pm	159.39	$ & $	1,015	\pm	370.86	$ & $	51.80	\pm	20.44	$ \\
~ & Chrome	& $	706	\pm	75.22	$ & $	1,817	\pm	251.04	$ & $	39.80	\pm	23.03	$ \\
\hline													
\multirow{2}{*}{\painter} & Firefox	& $	257	\pm	49.64	$ & $	1,185	\pm	216.02	$ & $	15.80	\pm	6.65	$ \\
~ & Chrome	& $	11,653	\pm	2,130.17	$ & $	13,500	\pm	2,477.33	$ & $	352.20	\pm	443.88	$ \\
\hline													
\multirow{2}{*}{\domtris} & Firefox	& $	49	\pm	13.42	$ & $	115	\pm	17.21	$ & $	11.20	\pm	2.17	$ \\
~ & Chrome	& $	117	\pm	12.45	$ & $	181	\pm	45.04	$ & $	7.40	\pm	2.19	$ \\
\hline													
\multirow{2}{*}{\colorgame} & Firefox	& $	352	\pm	77.53	$ & $	708	\pm	62.95	$ & $	6.00	\pm	3.87	$ \\
~ & Chrome	& $	210	\pm	40.08	$ & $	460	\pm	7.26	$ & $	5.60	\pm	2.61	$ \\
        \hline
    \end{tabular}
    \label{tab:update}
\end{table}

\subsubsection{Time taken to promote follower}
\label{sec:promote}

This experiment measures how long it takes to promote the follower to be the new leader (and demote the leader to become a follower) once the follower is up-to-date (\ie after the follower processes all events sent by the coordinator).
The experiment uses two browsers:  $B_1$ as the initial leader, and $B_2$ as the initial follower.
We execute half the workload by interacting with $B_1$, then switch their roles, then finish the workload by interacting with $B_2$.
We checked visually that the experiment behaves as expected, and measure the time taken since pressing the \emph{``Switch roles''} button until the roles are actually switched.
Column ``Promote'' on Table~\ref{tab:update} shows the results.
We can see that all promotions happen well under $100ms$, except for \painter on Google Chrome, which takes $352ms$.

\subsubsection{Time to perform an update}

Putting together Sections~\ref{sec:catchup} and~\ref{sec:promote} allows us to estimate the minimum time required to perform an update.
Even though it may take a follower browser as long as 13 seconds to synchronize its state with the leader, this process takes place in the background and does not cause the user to stop interacting with the (leader) browser.
Then, once the user is ready to switch browsers, the promote/demote process takes $352ms$, which humans perceive as instantaneous.
These two experiments also allow us to answer \textbf{RQ4}: \textbf{\sysname requires an imperceptible pause (353ms) to update a running browser, and requires 13 seconds (at most) to prepare that update in the background since launching the updated browser.}

\subsection{Using \sysname as an \mvx system}
\label{sec:mvxtime}

At its core, \sysname is an \mvx system targeting \js.
In this role, we are interested in measuring the latency between an event being triggered on the leader, and that same event being visible on the follower.
We designed an experiment that measures the \emph{Round-Trip Time (RTT)} of each event by
sending an acknowledgement from the follower, for each event received, back to the leader, through the coordinator.
The RTT provides a reasonable estimate of the leader-follower latency.
This experiment runs the workloads for all the applications while measuring the RTT.
Table~\ref{tab:rtt} shows the results.

In all cases, the RTT is under 110ms, which indicates a leader-follower latency of half the RTT, around 55ms.
The results from this experiment can answer \textbf{RQ5}: \textbf{Used as an \mvx system, \sysname delivers events to the follower about 55ms after the leader received them.}

\begin{table}[t]
    \centering
    \caption{Round-Trip Time (RTT) between the leader triggering an event and receiving an acknowledgement from the follower for that event.  Average of 5 runs with standard deviation.}
    \begin{tabular}{| l | l | c |}
        \hline
        \textbf{Program} & \textbf{Browser} & \textbf{Round-Trip Time (ms)} \\
        \hline

\multirow{2}{*}{\nicedit} & Firefox	& $	110.22	\pm	24.75	$ \\
~ & Chrome	& $	84.40	\pm	14.57	$ \\
\hline					
\multirow{2}{*}{\painter} & Firefox	& $	9.95	\pm	1.20	$ \\
~ & Chrome	& $	16.72	\pm	0.44	$ \\
\hline					
\multirow{2}{*}{\domtris} & Firefox	& $	10.05	\pm	0.34	$ \\
~ & Chrome	& $	6.58	\pm	0.16	$ \\
\hline					
\multirow{2}{*}{\colorgame} & Firefox	& $	28.26	\pm	3.45	$ \\
~ & Chrome	& $	25.45	\pm	1.23	$ \\

        \hline
    \end{tabular}
    \label{tab:rtt}
\end{table}

\subsection{Limitations and Untested Features}

The main design goal of \sysname is to allow instantaneous and stateful browser updates.
As such, we designed \sysname under the assumption that only one user interacts with each \js program, and that each \js program does not execute for a long time.
All these assumptions break for server-side \js applications written in node.js~\cite{nodejs}:  many users interact with each \js program, and each program executes for a long time.
Even though \sysname can be applied to such programs, to update the node.js virtual machine, this is not feasible, as such applications handler numerous events within a short time span and result in very large log files.
This is outside of the scope of \sysname.

\sysname has leader-specific and follower-specific code, that only executes in one browser.
Such code cannot call \texttt{document.createElement} or any of the other \js functions that \sysname intercepts.
Doing so creates a divergence that breaks \mvx under \sysname.
During development, we found that jQuery's \texttt{(eventObject.target).attr("id")} calls \texttt{document.createElement} internally (during internal unit testing).
Luckily, we were able to implement \sysname's prototype without using such functions.
In the future, as \sysname's functionally grows, we may need to ship our own modified versions of certain libraries.

\sysname cannot handle time-sensitive HTML5 elements, such as the \texttt{video} element~\cite{video} which embeds a media player to support video playback.
We note that \sysname supports stateful HTML5 elements, such as \texttt{canvas} elements~\cite{canvas} used by \painter in the evaluation.
However, \sysname needs to track time between events on the \texttt{video} element to ensure a correct replay (\eg play the video for 20 seconds, then pause it at the 20 second mark).
This is not supported by the current prototype, but we believe \sysname can be extended to support this feature.

\section{Related Work}

The problem of Dynamic Software Updating (\dsu) has been a focus of past research, resulting in \dsu systems for programs written in popular languages such as C~\cite{polus,kitsune,10.5555/2717477.2717485,10.1145/2663165.2663328,10.1145/2490301.2451147} and Java~\cite{rubah,jvolve,dcevm,dusc,jrebel,dustm}.
\sysname differs from these systems in two important ways.
First, \sysname updates the \emph{execution environment} and not the program running on that environment.
For instance, \dsu systems for Java do not support updating the underlying Java Virtual Machine while running the same program, which would be the closest to the goal of \sysname.
In fact, to the best of our knowledge, \sysname is the first such \dsu system outside of the Smalltalk community~\cite{smalltalk,pharo} to target the execution environment specifically.
Second, \dsu systems typically require modifications to the programs being updated to support stopping the program in one version and resuming it in the next, and to express how to transform the state in the old program to an equivalent representation that is compatible with the updated code.
In contrast, \sysname works on unmodified closed-source commercial internet browsers.
Instead of migrating the state directly, \sysname launches the new browser as a separate process and migrates the state for each page individually.
The only state kept outside of \sysname is persistent HTTP connections, which \sysname's proxy keeps open during updates.

\sysname uses Multi-Version eXecution (\mvx) to synchronize the old and new versions of the updated browser.
\mvx has been used mostly in programs written in C/C++ by intercepting and synchronizing system-calls between processes.
The main goal of \mvx are: (1) to increase security~\cite{mvarmor,nvxframework,remon}, detecting divergences in potentially suspect processes; (2) to increase reliability~\cite{varan,diehard,532621,tachyon,orchestra}, tolerating faults in one process by using the other processes; and (3) availability~\cite{mx,mvedsua,muc}, by performing updates on a forked process and terminating it when updates fail, without any disruption.
In fact, Mvedsua~\cite{mvedsua} is the most similar \mvx system to \sysname, given that it also combines \mvx with \dsu and allows users to build confidence on the validity of the update by executing both old and new versions for a period of time.
However, Mvedsua targets C programs updated via Kitsune~\cite{kitsune}, intercepts system calls, requires modifications to the programs being updated and machine-parseable descriptions of the update-induced divergences.
\sysname requires much less developer effort, which can be fully automated using an HTTP proxy.

Record-Replay (RR) can be described as ``offline Multi-Version eXecution''.
It operates in two phases, typically using two different (automatically generated) programs.
First, it records all non-determinism observed during an execution in a log file.
Then, it uses that log file to replay the same execution over the same program.
By contrast, \mvx records each non-deterministic datum in one process and replays it immediately on another process, thus keeping the state on both processes perfectly synchronized.
\mvx also needs to account for differences in execution speed that may result in a replayer overtaking the recorder and reaching a program point that requires non-determinism before that data is available.
For this reason, RR approaches require the log to be complete before being able to replay it.
Furthermore, RR approaches do not allow a replayer to become a recorder as they target a different problem: Accurate replication of bugs observed in production during development.

Techniques for RR target programs written in multiple popular programming languages:  C~\cite{rr}, Java~\cite{chroniclerj}, and even web browsers~\cite{mugshot,dolos,jardis,xcheck,rejs}.
Most techniques require a modified web browser.
Dolos~\cite{dolos} and Jardis~\cite{jardis} use modified implementations of browser components (Webkit and ChakraCore) to record bugs in production and replay them in development and provide developers with time-travel debugging capabilities, respectively.
Jardis focuses on node.js applications~\cite{nodejs}.
ReJS~\cite{rejs} also provides support for time-travelling debugging, but for any \js code in general, through a modified version of Microsoft's ChakraCore \js engine that performs heap checkpoints via a modified garbage-collector.

Working in pure \js, Mugshot~\cite{mugshot} is an RR system that demonstrates the feasibility of capturing all the needed non-determinism to ensure an accurate replay without the need for a modified web browser.
Mugshot influenced the design of \sysname by listing all the sources of non-determinism that need to be handled to capture all interactions between the environment and a \js program executing in a browser.
However, Mugshot relies on event listeners on the topmost DOM element (\ie \texttt{window}) to intercept all events, and replays them through synthetic browser events (\ie \texttt{DOMElement.fireEvent}).
As a result, Mugshot has to deal with browser-specific behavior that impacts event bubbling and event ordering.
\sysname's approach of intercepting each handler individually avoids such complexity and naturally supports any browser without special handling.
Similarly to \sysname, X-Check~\cite{xcheck} also works in pure \js and works on different browsers (all other techniques require the same browser and version to replay the recorded logs).
X-Check records logs on one browser and replays them on different browsers, with the goal of detecting cross-browser differences that developers can then replicate and address.

The closest system to \sysname is Cocktail~\cite{cocktail}, an \mvx system for web browsers with the goal of improving the security of internet browsers by feeding input to many different browsers and voting on the output.
Cocktail can thus detect and defeat attacks that target a particular browser, or a particular browser version.
Despite the very different goal, there are more important differences between \sysname and Cocktail.
First, Cocktail is implemented as a browser plugin and \sysname is implemented in pure \js.
As such, \sysname can be directly applied to any web browser, but Cocktail requires developer effort to write a new plugin for that browser.
Also, Cocktail's plugin can intercept synchronous non-determinism, such as calls to \texttt{Math.random}, and block until all browsers reach the same point.
This is not possible in \js's execution model, as described in Section~\ref{sec:eventloop}.
Second, Cocktail relies on an UI component to intercept mouse and keyboard events before they reach the browser.
\sysname captures the events at a finer level of detail, ensuring that all browsers execute the same \js handlers by the same order, regardless of implementation-specific browser quirks that may show the same element on different positions in different websites.
In fact, \sysname can replicate the execution even if the leader and follower have different window dimensions, which is a limitation of Cocktail.

\section{Conclusions and Future Work}

This paper presented the design of \sysname, a system that allows to update internet browsers without losing any state in the process.
\sysname works fully at the \js level, using first-class function interception to keep track of all events, and then using \mvx to perform updates on the new version of the browser while the old version keeps providing service.
As a result, \sysname works on popular, closed-source, commercial internet browsers such as Google Chrome.
\sysname requires a small amount of \js source changes, performed to each page opened in the target browser.
The changes required are easy to automate with a sophisticated internet proxy.

This paper also presented an extensive experimental evaluation, where \sysname is applied to \js applications with different combinations of features.
When not performing an update, \sysname imposes low overhead on the execution of event handlers (a max increase of 17.3ms).
Also, the state that \sysname keeps to support future updates grows at a modest rate of 23.5KB/s (at most) during intense user interaction.

\sysname can perform updates in short order, requiring just 13.5s (at most) to transfer the state from the old browser to the new browser.
While \sysname transfer the state, the user keeps interacting with the old browser.
Then, to finish the update and allow the user to interact with the new browser version, \sysname requires a very short pause in user interaction of 353ms (at most), which is barely noticeable for most users.

Besides its role in browser updates, \sysname doubles as an \mvx tool for \js applications.
The experimental evaluation showed that \sysname can keep two browsers synchronized, with an action on one browser taking effect on the other almost instantaneously, within 55ms.

In the modern internet age, an up-to-date internet browser is the first line of user defense.
\sysname lowers that barrier by transferring all the state kept in open pages from the old to the new browser, effectively eliminating any data loss or service interruption due to browser updates.
We strongly believe that \sysname has the potential to improve the average user's safety by making browser updates a thing of the past.

\subsection{Future Work}

The proxy component of \sysname ensure automatic protection on all the pages opened by an internet browser.
However, we believe \sysname could be adopted by the browser itself, either as a plugin or as part of the browser's code.
This paper shows that the modifications required to do so are minimal (just move the role of the proxy to inside the browser).

It is possible to use \sysname to move a page from one browser to another (\eg from Mozilla Firefox to Google Chrome).
This feature can be valuable to security conscious users, who can switch browsers as a vulnerability is disclosed.
We tested this feature of \sysname informally to ensure it works, but did not evaluate it or develop it further.

Requiring the user to check two pages for discrepancies can be automated.
For instance, \sysname can take two screenshots and then use computer vision algorithms to compare them~\cite{xcheck, cocktail}.
Moreover, \sysname can do that over a period of time (\eg 10s) to improve confidence.

\sysname is OS and platform agnostic, and we plan to implement \sysname on popular platforms (\eg Microsoft Windows and Apple OSX) and apply \sysname to the official browser in each platform (\eg Microsoft Edge and Apple Safari).


\bibliographystyle{ACM-Reference-Format}
\bibliography{luis}


\end{document}